\documentclass[%
aip,
amsmath,amssymb,
reprint,%
]{revtex4-2}
\usepackage{graphicx}
\usepackage{dcolumn}
\usepackage{mathrsfs}
\usepackage{bm}
\usepackage{lineno}
\usepackage{hyperref}
\usepackage{bbm}
\usepackage{ulem}
\usepackage{xcolor}

\begin{document}

\title{Time-domain measurement of ultra-fast intensity difference squeezed pulse pairs generated in fiber}

\author{Wen Zhao}
\affiliation{College of Precision Instrument and Opto-Electronics Engineering, Key Laboratory of Opto-Electronics Information Technology, Ministry of Education, Tianjin University, Tianjin 300072, People’s Republic of China\\}
\author{Xueshi Guo}
 \email{xueshiguo@tju.edu.cn}
\affiliation{College of Precision Instrument and Opto-Electronics Engineering, Key Laboratory of Opto-Electronics Information Technology, Ministry of Education, Tianjin University, Tianjin 300072, People’s Republic of China\\}

\author{Xiaoying Li}
 \email{xiaoyingli@tju.edu.cn}
\affiliation{College of Precision Instrument and Opto-Electronics Engineering, Key Laboratory of Opto-Electronics Information Technology, Ministry of Education, Tianjin University, Tianjin 300072, People’s Republic of China\\}

\begin{abstract}
Pulsed pumped four-wave mixing process via $\chi^{(3)}$ non-linearity in optical fiber can generate optical pulses with continuous variable quantum correlation. However, pair-wise correlation of the generated pulses in this system has not been studied. Here we report a time-domain measurement of an intensity difference squeezed state generated in fiber. With a fast response differential detection system, we show the generated twin-beam pulses are pair-wisely correlated, and -3.8 dB (-8.1 dB after detection losses correction) intensity difference squeezing degree is measured in time-domain. Our result is benefit for generating multi-mode entangled state by time-division multiplexing in fiber system.
\end{abstract}

\maketitle

Continuous variables (CV) optical quantum state are useful resources for multiple quantum information schemes on aspects of quantum sensing, quantum communication and quantum computation \cite{Braunstein2005RevModPhys, Adesso2014OSID}. Ultra-fast pulse pumped $\chi^{(3)}$ non-linearity of Kerr or four-wave mixing process in optical fibers has been proved to be one of the good candidates for CV optical quantum state generating \cite{Schmitt1998PRL,sharping2001observation, guo2012APL,liu2018OL, Nannan2016OE, Silberhorn2001PRL, Guo2016OL, Li2019OE}. In different fiber based experimental system, making use of the long interaction length and the ultra-low loss property of fiber, people have successfully generate photon number squeezed solitons \cite{Schmitt1998PRL}, intensity difference squeezed twin-beams \cite{sharping2001observation, guo2012APL,liu2018OL}, single mode squeezed state \cite{Shelby1986PRL, Nannan2016OE} and CV Einstein-Podolsky-Rosen (EPR) entanglement \cite{Silberhorn2001PRL, Guo2016OL, Li2019OE}. Pumped by a mode-locked laser, the generated optical pulses can be viewed as CV quantum state copies of identical parameters with equal time interval. As it has been demonstrated by the recent work in continuous-wave (CW) or quasi-continuous wave (QCW) laser pumped optical parametric oscillator system, these states can be used to produce large scale CV multi-mode entanglement state by time-division multiplexing and coherent combination \cite{Yoshikawa2016APLPhotonics, Asavanant2019science, Larsen2019science}. The multiplexing window of an ultra-fast pulse pumped system defined by the time interval between two adjacent pulses of the mode-locked laser is of 10 ns level, which is obviously short than that in current reported CW or QCW system. This short time window property will help to increase the capacity of the entanglement state. Besides, quantum state generated by ultra-short pulse pumped system contains multiple temporal and frequency mode \cite{HuoNan2020PRL}, which can serve as an extra degree of freedom in the entanglement state generation process.

So far, the existing CV state generated in fiber systems are all analyzed in frequency-domain, and the result tells the average quantum statistical property of many copies of the quantum states resident in optical pulses. Refs. \cite{shinjo2019pulse, OkuboOL08} demonstrated an ultra-short optical pulse train of CV EPR state generated from $\chi^{(2)}$ nonlinear crystal and measured with time-domain method, but similar measurement for pulsed intensity difference squeezing has not been reported. In this letter, we report a time-domain measurement of an intensity difference squeezed state generated in an ultra-fast pulse pumped fiber optical parametric amplifier (FOPA). The optical transmitting efficiencies for the frequency non-degenerate signal and idler outputs of the FOPA are about 97$\%$ and 95$\%$, respectively. By coupling the signal and idler pulse pairs onto a fast response differential detector with a quantum efficiency of about 72$\%$, we show the intensity difference squeezing (IDS) degree measured with time-domain method is -3.8 dB (-8.1 dB after detection losses correction). By analyzing the voltage statistical properties of the differential detector output, we show our time domain method can efficiently reveal the pairwise quantum correlation between the signal and the idler pulses generated from the FOPA. We further discuss possible ways to improve IDS degree by performing high efficiency frequency-domain measurement. The result shows a record high directly measured IDS degree of -7.2 dB (-10.1 dB after detection losses correction) in fiber system \cite{liu2018OL}, which also implies the FOPA system has the potential to achieve -10 dB IDS degree if the detection system is further optimized.

\begin{figure}[h]
	\centering
	\includegraphics[width=0.95\linewidth]{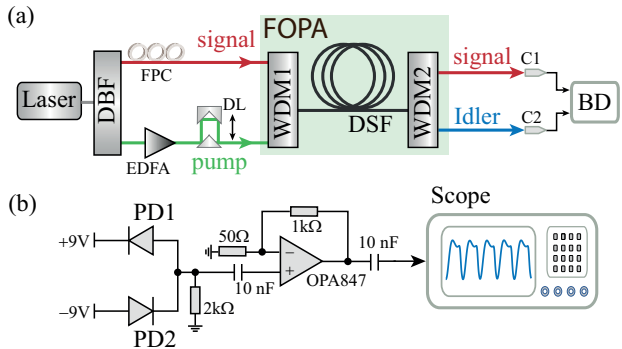}
	\caption{(a) The experimental setup;  (b) The simplified electronic schematics for balanced detector (BD) system. FOPA, fiber optical parametric amplifier; DBF, dual-band filter; WDM, wavelength division-multiplexer; EDFA, erbium-doped fiber amplifier; FPC, fiber polarization controller; DL, delay line; DSF, dispersion-shifted fiber; C1-C2, fiber collimator; PD1-PD2, photo-diode. 
	}
	\label{IDS-steup-PSD}
\end{figure}

Our experimental setup is shown in Fig.~\ref{IDS-steup-PSD}(a), where the FOPA consists of a 300 m dispersion shifted fiber (DSF) and two wavelength division-multiplexers (WDM1, WDM2). The pump source for the FOPA is a mode-locked fiber laser (MenloSystems LAC1550) with a central wavelength of about 1550 nm and a full width at half maximum (FWHM) of about 100 nm. The pulse repetition rate of the laser is around 50 MHz, which defines the time-window of the generated quantum state. Pico-second pump pulses and the injected signal pulses are obtained by filtering the mode-locked fiber laser through a tunable dual-band filter (DBF) realized by optical gratings. The output spectra of the DBF for both the pump and signal are Gaussian shaped, the FWHM of the spectra for the pump and signal are about 0.9 nm and 1.5 nm, respectively. An erbium-doped fiber amplifier (EDFA) is used to amplify the pump pulses to achieve the required pump power. The zero dispersion wavelength of the DSF is about 1548.5 nm, and the center wavelengths for the pump and the injected signal are respectively set to 1549.3 nm and 1533.0 nm to ensure the satisfaction of the phase matching condition for the four-wave mixing process. The polarization and the temporal modes of the pump and the signal are well matched by using the fiber polarization controller (FPC) and delay line (DL). The mode matched pump and the signal are combined by WDM1 and send to the 300 m DSF, which functions as the gain media of the FOPA. In order to adjust the intensity and spectra of both the pump and the signal fields, we use a bench top wave shaper (WS, Finisar 4000A) to realize WDM1. In the DSF, the injected signal is amplified by the pump and the pulsed idler field with a spectrum centered at 1566.0 nm is generated. The signal and idler pulses are spatially separated with a two channel coarse wavelength  division-multiplexer (WDM2). The optical setup is similar to our previous work reported in \cite{liu2018OL}. However, the transmission efficiencies for the signal channel and the idler channel are improved to about 97$\%$ and 95$\%$ by optimizing the optical path and using high efficiency coarse WDM filters. The signal and the idler pulses output from WDM2 are sent to a fast response balanced detector (BD) system through two fiber collimators (C1, C2). 

The simplified electronic schematics of the BD is shown in Fig.~\ref{IDS-steup-PSD}(b). Two fiber pig-tailed photo diodes (PD1 and PD2, SWT PDS123-CFA-B0202) are reversely biased with $\pm 9$ V and the anode of one PD is directly connect to the cathode of the other to convert the intensity difference of the signal and the idler fields into photo current. This differential photo-current is further convert to voltage by a 2 k$\Omega$ load, amplified by an operation amplifier (TI OPA847) working in non-inverting gain of 21 V/V and recorded by a scope with a voltage resolution of 12 bits. The quantum efficiency of the BD, including fiber coupling efficiency, is about 72$\%$, so the overall detection efficiencies for the signal and the idler channel are about $70\%$ and $68\%$, respectively. Since the time duration of each optical pulse is much smaller than the response time of the BD, the recorded output voltage $v(t)$ from a well balanced BD is related to the intensity difference of the signal and the idler fields by \cite{Ou2017Book}
\begin{equation}
	\label{eq:DDS_out}
	v(t) = \sum_{n=1}^{N_t} k(t-n \Delta T )  I_{d,n},
\end{equation} 
where $N_t$ is the total number of the pulse pairs measured in a single data acquisition process, $k(\cdot)$ is the response function of the BD, $\Delta T$ is the time interval between two adjacent laser pulses and $I_{d,n} = I_{s,n} - r I_{i,n}$ is the intensity difference of signal ($I_{s,n}$) and the generated idler ($ I_{i,n}$) for the n$^{th}$ pulse pairs. We note $r$ is an electronic-gain-ratio parameter before photo current subtraction which only relies on the design of the BD \cite{guo2013PRA}, and we have $r\approx1$ in our experiment. To avoid saturating the operation amplifier used in the BD, we introduce a tunable loss to either the signal or the idler channel by tilting the coupling to C1 or C2, and make sure the averaged intensity difference always satisfy $  \bar I_{d,n} = \sum_{n=1}^{N_t} I_{d,n}/N_t \approx 0$. Eq.(\ref{eq:DDS_out}) shows the response overlapping from different optical pulses are negligible when the response function $k(\cdot)$ is fast enough. Therefore, the intensity difference estimation to individual pulse pairs of signal and idler can be obtained independently by simply analyzing $v(t)$ in different time window with the time duration of $\Delta T$.

\begin{figure*}[ht]
	\centering
	\includegraphics[width=0.95\linewidth]{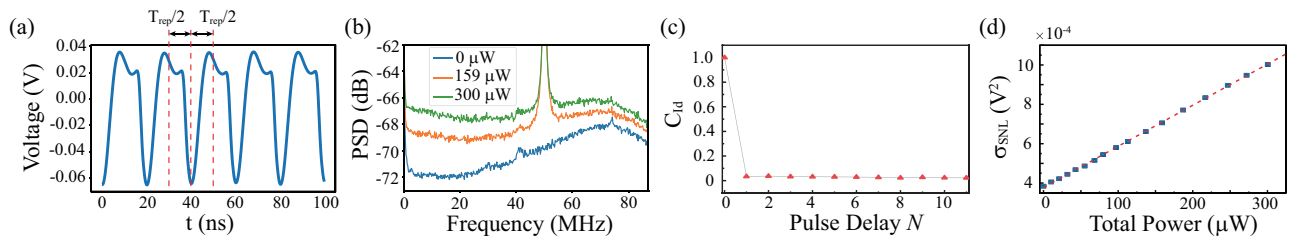}
	\caption{\label{BDcal}Calibration results for the fast response balanced detection system (BD). (a) Voltage data acquired by the digital oscilloscope over 5 pulse periods. (b) The shot noise power spectral density (PSD) with different optical input powers. (c) The normalized correlation coefficient $C_{Id}$ between the $n^{th}$ and $(n+N)^{th}$ pulse-voltage integral value measured with the fast BD in shot-noise calibration process.(d) Shot-noise of the BD calibrated under different optical input power. The dashed line is the linear fit to the data points.}
\end{figure*}

We first illustrate the calibration of the BD and our data acquisition (DAQ) procedure for time-domain measurement of the intensity difference fluctuation. The two light used for this calibration is obtained by directly filtering the laser to obtain a 0.9 nm Gaussian shaped spectrum centered at 1549.3 nm, and sending this filtered light to a 50:50 beam splitter. The BD has a high common mode rejection ratio (CMMR) of more than 50 dB, so the above mentioned setup can efficiently calibrate the shot noise limit (SNL) of the BD though the probe light itself is not shot noise limited. We note the wavelength dependence of the BD response is negligible within the bandwidth we are interested in. Fig.~\ref{BDcal}(a) shows the typical output waveform of the BD analyzed by the scope, which gives out a voltage pulse for each detected pulse pair. To characterize the statistics property of the pulse pairs sent to the BD, we repeat the DAQ process for 1000 times in the same experimental condition. For each single DAQ process, 25 k samples are recorded with the sampling rate of 5G sample/s. This corresponds to a time window of 5 $\mu$s and includes $N_t=250$ optical pulses. To estimate the bandwidth of the BD, we calculate the power spectral density (PSD) by doing fast Fourier transformation to the 25 k samples on a computer, and average over 1000 measurement times. The results for different light power are shown in Fig.~\ref{BDcal}(b). The peak at 50 MHz is due to the residue response to the laser repetition rate. The PSDs indicate the BD has a bandwidth of about 80 MHz. This is larger than the repetition rate of the  mode-locked laser, which is a essential condition to distinguish each pulsed quantum state in time domain measurement \cite{OkuboOL08}.

To estimate the intensity difference of individual pulse pairs, we find the negative peak aroused by each optical pulse pair and sum the acquired data points within $\pm \Delta T/2$, as shown by the red dashed line in Fig.\ref{BDcal}(a). Then the summation result $e_{n}$ is used as the estimation value of intensity difference between the $n^{th}$ optical pulse pairs send to the BD at the same time. In our DAQ process, $n^{th}$ takes positive integer value smaller than $N_t$. To verify the response function of the fast BD can distinguish each pulse pair independently, we estimate the normalized correlation coefficient between the calculated $e_n$ values and its $N$-pulses-shifted values $e_{n-N}$ as it is analyzed in Refs. \cite{LiYongmin2018JOSAB, shinjo2019pulse}
\begin{equation}
	\label{eq:corr_coef}
	C_{Id}(N) = \frac{1}{N_t-N} \sum_{n=N}^{N_t}  \frac{ ( e_{n} - \bar e) \cdot (e_{n-N} - \bar e )} { \sigma^2 },  
\end{equation}
where $N$ is the shifting index, $\bar e $ is the expect value of $e_n$, and $\sigma^2 $ is the variance of $e_n$. Eq.(2) obviously has the property of $C_{Id}(0)=1$ for $N=0$. The calculated $C_{Id}(N)$ at different $N$ for our BD are shown by the triangles in Fig.~\ref{BDcal}(c), which is obtained by averaging over 1000 DAQ process for the same experimental condition. We find $C_{Id}(N)$ drops to below $6\%$ for just one pulse delay and the value of $C_{Id}(N)$ varies small but randomly for $N=1$ to 10. We think this residue correlation may come from either the drifting of BD balancing or electronic coupling from the environment. This result indicates our fast BD can measure the optical pulses almost independently. With the above mentioned time-domain DAQ method, we calibrate the SNL of the BD in different optical power. The calibrated variance $\sigma_{SNL}$ is shown with blue squares in Fig.~\ref{BDcal}(d), and the dashed line is the linear fit of the squares, which shows good linearity between the optical power and the variance of the voltage difference integral value. This indicates the extra noise of the calibration light is canceled out by the CMMR of the BD, and the electronic response of the BD has good linearity. 

\begin{figure}[ht]
\centering
\includegraphics[width=0.95\linewidth]{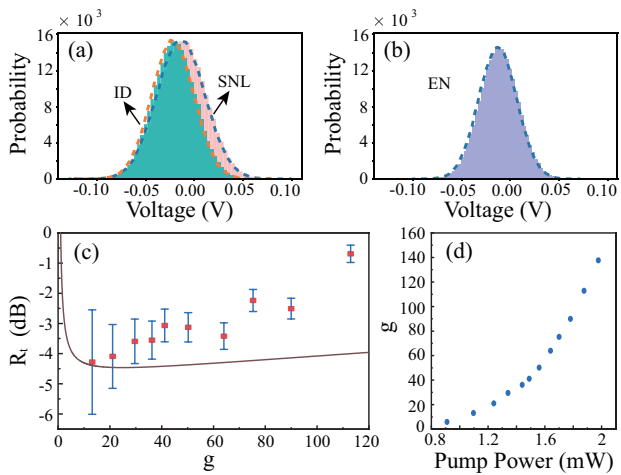}
\caption{(a) A typical histogram of integrated result $e_n$ for signal and idler pulse pairs (ID) and the corresponding shot noise (SNL). (b) Histogram of $e_n$ for the electronic noise of the BD. (c) The degree of intensity difference squeezing $R_t$ measured in time domain. The solid curve is the theoretical prediction of $R_t$ for detection efficiencies $\eta_s = 70\%$ and $\eta_i = 68\%$. (d) The power gain of the amplifier $g$ versus the pump power when the power of the injected signal is $I_s^{in}=1$ $\mu$W.
}
\label{TD-IDS}
\end{figure}

We use this fitting result in Fig.~\ref{BDcal}(d) as the benchmark to get the IDS degree. Compared to the previous work \cite{OkuboOL08} on time-domian measurement of continuous variable entanglement, the optical power sent to the BD is much smaller in an intensity difference squeezing measurement since no strong local oscillator light is used. So the contribution from the electronic noise of the BD to $e_n$ is not negligible in the DAQ process. Since this electronic noise is independent of the optical noise, we correct the influence of the electronic noise by subtracting the variance from both the variance of intensity noise and the SNL, and thus the IDS degree $R_t$ is experimentally estimated with
\begin{equation}
	\label{Rt}
	R_t  =  \frac{\sigma_{ID}^2  - \sigma_{EN}^2 }{\sigma_{SNL}^2  - \sigma_{EN}^2},
\end{equation} 
where $\sigma_{ID}^2$ is the variance of integrated value for signal and idler pulses measured by the BD, $\sigma_{SNL}^2$ is the SNL variance of the same total optical power calculated by the fitting parameters, and $\sigma_{EN}^2$ is the electronic noise induced variance on $e_n$ of the BD system. Fig.~\ref{TD-IDS}(a) shows a typical histogram of integrated result $e_n$ for 250 k pulse pairs where the power launched onto each PD is $P\approx 55$ $\mu$W.
The power gain of the amplifier $g=I_s^{out}/I_s^{in}$, defined as the ratio of the power for amplified signal  $I_s^{out}$ and the power for injected signal  $I_s^{in}$, is about 64, and the power for injected signal is $I_s^{in}\approx 1$ $\mu$W.
From this result, the variance of voltage integral value for the signal and idler fields and the corresponding SNL are estimated to be $\sigma^2_{\mathrm{ID}}\approx4.80 \times 10^{-4}~\mathrm{V^2}$ and $\sigma^2_{\mathrm{SNL}}\approx6.11 \times 10^{-4}~ \mathrm{V^2}$, respectively. 
We also show the probability distribution of voltage difference integral value of electronic noise of the measurement system without light injection in Fig.~\ref{TD-IDS}(b), and the fitted variance is $\sigma^2_{\mathrm{EN}}\approx3.86 \times 10^{-4}~ \mathrm{V^2}$. By using Eq.(\ref{Rt}), we find the measured quantum noise reduction is about $-3.80$ dB.

With the above mentioned experimental procedure and DAQ condition, we measure the IDS degree $R_t$ in different $g$, which can be controlled by varying the pump power of FOPA. Squares in Fig.~\ref{TD-IDS}(c) show the measured $R_t$ as a function of $g$, and the corresponding $g$ in different pump power are shown in Fig.~\ref{TD-IDS}(d). In this measurement, $R_t$ is about $-3.5$ dB when $g$ is within the range of 20-60. Using the experimental parameters and the theoretical model in Ref. \cite{guo2013PRA}, we calculate $R_t$ as a function of $g$ and the result is represented by the solid curve of Fig.~\ref{TD-IDS}(c). One sees $R_t$ for both the experimental data and the theoretical curve becomes worse as $g$ increases for $g>20$, but the experimentally measured $R_t$ is even worse than that of theoretically calculated result. We think this deviation between measurement and theory prediction is caused by the following reasons. 1) The classical noise of injected signal. This kind of noise can be amplified by the pump \cite{Larsen2019njpQI}, so its impairment becomes more obvious as $g$ increases. 2) The parasitic Raman scattering process in the amplification process. We note this effect can be reduced by further cooling the fiber \cite{2K_DSF}. 3) The pump depletion since the average power of the pulsed pump is comparable to that for the amplified signal and the idler light. 4) The parasitic higher-order four-wave mixing effect \cite{guo2012APL}. In principle, the impairment for these factors can be alleviate if the power of the injected signal is further reduced, but it is challenging to realize a fast response BD that are suitable for weak signal detection.
\begin{figure}[ht]
\centering
\includegraphics[width=0.83\linewidth]{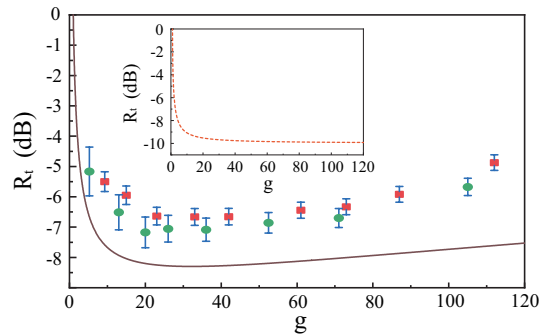}
\caption{\label{fre-IDS}The degree of intensity difference squeezing $R_t$ measured in frequency domain around 2.5 MHz. The results represented by circles and squares are obtained with signal injection $I_s^{in}$=0.2, 1 $\mu$W, respectively, and the solid curve is the theoretical prediction of $R_t$ for for detection efficiencies $\eta_s = 91\%$ and $\eta_i = 89\%$. The inset plots the theoretical prediction of $R_t$ with the same channel efficiencies but the electronic gain ratio of the BD is optimized according to Eq.(18) in Ref. \cite{guo2013PRA}.
}
\end{figure}

Despite of these factors, the main constrain of our time domain measurement is the detection efficiency of the BD since we are not able to implement a fast BD with high detection efficiency yet. To prove this, we also preform a frequency-domain measurement \cite{liu2018OL} of IDS around 2.5 MHz. In the measurement, we keep the experimental condition the same as the time domain measurement, but only replace the fast BD with a 20-MHz-bandwidth BD having detection efficiency of about $94\%$ leading to the overall detection efficiencies $\eta_s \approx 91\%$ and $\eta_i \approx 89\%$. The measured $R_t$ (squares) together with the theoretical calculated $R_t$ in different $g$ (solid curve) are shown in Fig.~\ref{fre-IDS}, from which one sees the measured IDS degree is improved to 6.7 dB. Therefore, we think the IDC degree measured in time domain can be improved  if the detection efficiency of the fast BD is increased. We then reduce the injection power to $I_s^{in} = 0.2~\mu W$, and the measured result is shown by circles in Fig.~\ref{fre-IDS}. The measured IDS is further improved to about $-7.2$ dB  (-10.1 dB after detection losses correction) for $g\approx20$, which is a record high squeezing degree in fiber system. Finally, we note that in both the time/frequency domain measurement, we realize $ I_{d,n} = I_{s,n} - r I_{i,n} \approx 0$ by fixing the coefficient $r=1$ and adding a tunable optical loss before the sensitive area of the photo diode. As it has been discussed in \cite{guo2013PRA}, this extra loss can be avoid if one can realize a variable electronic gain ratio $r$ before the subtraction of the photo current in the electronic design of the BD, and optimize the gain ratio $r$ during measurement. To illustrate this, we calculate the theoretical prediction of $R_t$ when $r$ is optimized and the channel efficiencies are set to $\eta_s = 91\%$ and $\eta_i = 89\%$. The result is shown in the inset of Fig.~\ref{fre-IDS}, which shows improvement compared to the solid curve in Fig.~\ref{fre-IDS} and gives a near -10 dB IDS degree when $g$ is large enough. 

In conclusion, we experimentally demonstrate a time-domain measurement of intensity difference squeezed optical pulse pairs generated by a fiber optical parametric amplifier. With a fiber coupled fast response BD that can distinguish each pulse, we show the IDS property happens independently to each signal idler optical pulse pairs. The intensity-difference noise of the twin beams drops below the SNL by $-3.8$ dB (-8.1 dB after detection losses correction). Theoretical analysis shows the IDS degree can be further optimized to -10 dB if the injected signal is small enough, Raman scatter is effectively restrained and the electronic gain of the BD is tunable. The result serves as a preliminary experimental test for generating multi-mode entangled state by using time-division multiplexing of the pulse repetition period in fiber system.

This work was supported in part by National Natural Science Foundation of China (Grants No. 12004279).

The authors declare no conflicts of interest.

Data underlying the results presented in this paper are not publicly available at this time but may be obtained from the authors upon reasonable request.

\bibliography{main}

\begin{thebibliography}{22}%
\makeatletter
\providecommand \@ifxundefined [1]{%
 \@ifx{#1\undefined}
}%
\providecommand \@ifnum [1]{%
 \ifnum #1\expandafter \@firstoftwo
 \else \expandafter \@secondoftwo
 \fi
}%
\providecommand \@ifx [1]{%
 \ifx #1\expandafter \@firstoftwo
 \else \expandafter \@secondoftwo
 \fi
}%
\providecommand \natexlab [1]{#1}%
\providecommand \enquote  [1]{``#1''}%
\providecommand \bibnamefont  [1]{#1}%
\providecommand \bibfnamefont [1]{#1}%
\providecommand \citenamefont [1]{#1}%
\providecommand \href@noop [0]{\@secondoftwo}%
\providecommand \href [0]{\begingroup \@sanitize@url \@href}%
\providecommand \@href[1]{\@@startlink{#1}\@@href}%
\providecommand \@@href[1]{\endgroup#1\@@endlink}%
\providecommand \@sanitize@url [0]{\catcode `\\12\catcode `\$12\catcode
  `\&12\catcode `\#12\catcode `\^12\catcode `\_12\catcode `\%12\relax}%
\providecommand \@@startlink[1]{}%
\providecommand \@@endlink[0]{}%
\providecommand \url  [0]{\begingroup\@sanitize@url \@url }%
\providecommand \@url [1]{\endgroup\@href {#1}{\urlprefix }}%
\providecommand \urlprefix  [0]{URL }%
\providecommand \Eprint [0]{\href }%
\providecommand \doibase [0]{https://doi.org/}%
\providecommand \selectlanguage [0]{\@gobble}%
\providecommand \bibinfo  [0]{\@secondoftwo}%
\providecommand \bibfield  [0]{\@secondoftwo}%
\providecommand \translation [1]{[#1]}%
\providecommand \BibitemOpen [0]{}%
\providecommand \bibitemStop [0]{}%
\providecommand \bibitemNoStop [0]{.\EOS\space}%
\providecommand \EOS [0]{\spacefactor3000\relax}%
\providecommand \BibitemShut  [1]{\csname bibitem#1\endcsname}%
\let\auto@bib@innerbib\@empty
\bibitem [{\citenamefont {Braunstein}\ and\ \citenamefont {van
  Loock}(2005)}]{Braunstein2005RevModPhys}%
  \BibitemOpen
  \bibfield  {author} {\bibinfo {author} {\bibfnamefont {S.~L.}\ \bibnamefont
  {Braunstein}}\ and\ \bibinfo {author} {\bibfnamefont {P.}~\bibnamefont {van
  Loock}},\ }\bibfield  {title} {\enquote {\bibinfo {title} {Quantum
  information with continuous variables},}\ }\href@noop {} {\bibfield
  {journal} {\bibinfo  {journal} {Rev. Mod. Phys.}\ }\textbf {\bibinfo {volume}
  {77}},\ \bibinfo {pages} {513--577} (\bibinfo {year} {2005})}\BibitemShut
  {NoStop}%
\bibitem [{\citenamefont {Adesso}, \citenamefont {Ragy},\ and\ \citenamefont
  {Lee}(2014)}]{Adesso2014OSID}%
  \BibitemOpen
  \bibfield  {author} {\bibinfo {author} {\bibfnamefont {G.}~\bibnamefont
  {Adesso}}, \bibinfo {author} {\bibfnamefont {S.}~\bibnamefont {Ragy}},\ and\
  \bibinfo {author} {\bibfnamefont {A.~R.}\ \bibnamefont {Lee}},\ }\bibfield
  {title} {\enquote {\bibinfo {title} {Continuous variable quantum information:
  Gaussian states and beyond},}\ }\href@noop {} {\bibfield  {journal} {\bibinfo
   {journal} {Open Systems \& Information Dynamics}\ }\textbf {\bibinfo
  {volume} {21}},\ \bibinfo {pages} {1440001} (\bibinfo {year}
  {2014})}\BibitemShut {NoStop}%
\bibitem [{\citenamefont {Schmitt}\ \emph {et~al.}(1998)\citenamefont
  {Schmitt}, \citenamefont {Ficker}, \citenamefont {Wolff}, \citenamefont
  {K\"onig}, \citenamefont {Sizmann},\ and\ \citenamefont
  {Leuchs}}]{Schmitt1998PRL}%
  \BibitemOpen
  \bibfield  {author} {\bibinfo {author} {\bibfnamefont {S.}~\bibnamefont
  {Schmitt}}, \bibinfo {author} {\bibfnamefont {J.}~\bibnamefont {Ficker}},
  \bibinfo {author} {\bibfnamefont {M.}~\bibnamefont {Wolff}}, \bibinfo
  {author} {\bibfnamefont {F.}~\bibnamefont {K\"onig}}, \bibinfo {author}
  {\bibfnamefont {A.}~\bibnamefont {Sizmann}},\ and\ \bibinfo {author}
  {\bibfnamefont {G.}~\bibnamefont {Leuchs}},\ }\bibfield  {title} {\enquote
  {\bibinfo {title} {Photon-number squeezed solitons from an asymmetric
  fiber-optic sagnac interferometer},}\ }\href@noop {} {\bibfield  {journal}
  {\bibinfo  {journal} {Phys. Rev. Lett.}\ }\textbf {\bibinfo {volume} {81}},\
  \bibinfo {pages} {2446--2449} (\bibinfo {year} {1998})}\BibitemShut {NoStop}%
\bibitem [{\citenamefont {Sharping}, \citenamefont {Fiorentino},\ and\
  \citenamefont {Kumar}(2001)}]{sharping2001observation}%
  \BibitemOpen
  \bibfield  {author} {\bibinfo {author} {\bibfnamefont {J.~E.}\ \bibnamefont
  {Sharping}}, \bibinfo {author} {\bibfnamefont {M.}~\bibnamefont
  {Fiorentino}},\ and\ \bibinfo {author} {\bibfnamefont {P.}~\bibnamefont
  {Kumar}},\ }\bibfield  {title} {\enquote {\bibinfo {title} {Observation of
  twin-beam-type quantum correlation in optical fiber},}\ }\href@noop {}
  {\bibfield  {journal} {\bibinfo  {journal} {Optics letters}\ }\textbf
  {\bibinfo {volume} {26}},\ \bibinfo {pages} {367--369} (\bibinfo {year}
  {2001})}\BibitemShut {NoStop}%
\bibitem [{\citenamefont {Guo}\ \emph {et~al.}(2012)\citenamefont {Guo},
  \citenamefont {Li}, \citenamefont {Liu}, \citenamefont {Yang},\ and\
  \citenamefont {Ou}}]{guo2012APL}%
  \BibitemOpen
  \bibfield  {author} {\bibinfo {author} {\bibfnamefont {X.}~\bibnamefont
  {Guo}}, \bibinfo {author} {\bibfnamefont {X.}~\bibnamefont {Li}}, \bibinfo
  {author} {\bibfnamefont {N.}~\bibnamefont {Liu}}, \bibinfo {author}
  {\bibfnamefont {L.}~\bibnamefont {Yang}},\ and\ \bibinfo {author}
  {\bibfnamefont {Z.~Y.}\ \bibnamefont {Ou}},\ }\bibfield  {title} {\enquote
  {\bibinfo {title} {An all-fiber source of pulsed twin beams for quantum
  communication},}\ }\href@noop {} {\bibfield  {journal} {\bibinfo  {journal}
  {Applied Physics Letters}\ }\textbf {\bibinfo {volume} {101}},\ \bibinfo
  {pages} {261111} (\bibinfo {year} {2012})}\BibitemShut {NoStop}%
\bibitem [{\citenamefont {Liu}\ \emph {et~al.}(2018)\citenamefont {Liu},
  \citenamefont {Huo}, \citenamefont {Li},\ and\ \citenamefont
  {Li}}]{liu2018OL}%
  \BibitemOpen
  \bibfield  {author} {\bibinfo {author} {\bibfnamefont {Y.}~\bibnamefont
  {Liu}}, \bibinfo {author} {\bibfnamefont {N.}~\bibnamefont {Huo}}, \bibinfo
  {author} {\bibfnamefont {J.}~\bibnamefont {Li}},\ and\ \bibinfo {author}
  {\bibfnamefont {X.}~\bibnamefont {Li}},\ }\bibfield  {title} {\enquote
  {\bibinfo {title} {Long-distance distribution of the telecom band intensity
  difference squeezing generated in a fiber optical parametric amplifier},}\
  }\href@noop {} {\bibfield  {journal} {\bibinfo  {journal} {Optics Letters}\
  }\textbf {\bibinfo {volume} {43}},\ \bibinfo {pages} {5559--5562} (\bibinfo
  {year} {2018})}\BibitemShut {NoStop}%
\bibitem [{\citenamefont {Liu}\ \emph {et~al.}(2016)\citenamefont {Liu},
  \citenamefont {Liu}, \citenamefont {Li}, \citenamefont {Yang},\ and\
  \citenamefont {Li}}]{Nannan2016OE}%
  \BibitemOpen
  \bibfield  {author} {\bibinfo {author} {\bibfnamefont {N.}~\bibnamefont
  {Liu}}, \bibinfo {author} {\bibfnamefont {Y.}~\bibnamefont {Liu}}, \bibinfo
  {author} {\bibfnamefont {J.}~\bibnamefont {Li}}, \bibinfo {author}
  {\bibfnamefont {L.}~\bibnamefont {Yang}},\ and\ \bibinfo {author}
  {\bibfnamefont {X.}~\bibnamefont {Li}},\ }\bibfield  {title} {\enquote
  {\bibinfo {title} {Generation of multi-mode squeezed vacuum using pulse
  pumped fiber optical parametric amplifiers},}\ }\href@noop {} {\bibfield
  {journal} {\bibinfo  {journal} {Opt. Express}\ }\textbf {\bibinfo {volume}
  {24}},\ \bibinfo {pages} {2125--2133} (\bibinfo {year} {2016})}\BibitemShut
  {NoStop}%
\bibitem [{\citenamefont {Silberhorn}\ \emph {et~al.}(2001)\citenamefont
  {Silberhorn}, \citenamefont {Lam}, \citenamefont {Wei\ss{}}, \citenamefont
  {K\"onig}, \citenamefont {Korolkova},\ and\ \citenamefont
  {Leuchs}}]{Silberhorn2001PRL}%
  \BibitemOpen
  \bibfield  {author} {\bibinfo {author} {\bibfnamefont {C.}~\bibnamefont
  {Silberhorn}}, \bibinfo {author} {\bibfnamefont {P.~K.}\ \bibnamefont {Lam}},
  \bibinfo {author} {\bibfnamefont {O.}~\bibnamefont {Wei\ss{}}}, \bibinfo
  {author} {\bibfnamefont {F.}~\bibnamefont {K\"onig}}, \bibinfo {author}
  {\bibfnamefont {N.}~\bibnamefont {Korolkova}},\ and\ \bibinfo {author}
  {\bibfnamefont {G.}~\bibnamefont {Leuchs}},\ }\bibfield  {title} {\enquote
  {\bibinfo {title} {Generation of continuous variable einstein-podolsky-rosen
  entanglement via the kerr nonlinearity in an optical fiber},}\ }\href@noop {}
  {\bibfield  {journal} {\bibinfo  {journal} {Phys. Rev. Lett.}\ }\textbf
  {\bibinfo {volume} {86}},\ \bibinfo {pages} {4267--4270} (\bibinfo {year}
  {2001})}\BibitemShut {NoStop}%
\bibitem [{\citenamefont {Guo}\ \emph {et~al.}(2016)\citenamefont {Guo},
  \citenamefont {Liu}, \citenamefont {Liu}, \citenamefont {Li},\ and\
  \citenamefont {Ou}}]{Guo2016OL}%
  \BibitemOpen
  \bibfield  {author} {\bibinfo {author} {\bibfnamefont {X.}~\bibnamefont
  {Guo}}, \bibinfo {author} {\bibfnamefont {N.}~\bibnamefont {Liu}}, \bibinfo
  {author} {\bibfnamefont {Y.}~\bibnamefont {Liu}}, \bibinfo {author}
  {\bibfnamefont {X.}~\bibnamefont {Li}},\ and\ \bibinfo {author}
  {\bibfnamefont {Z.~Y.}\ \bibnamefont {Ou}},\ }\bibfield  {title} {\enquote
  {\bibinfo {title} {Generation of continuous variable quantum entanglement
  using a fiber optical parametric amplifier},}\ }\href@noop {} {\bibfield
  {journal} {\bibinfo  {journal} {Opt. Lett.}\ }\textbf {\bibinfo {volume}
  {41}},\ \bibinfo {pages} {653--656} (\bibinfo {year} {2016})}\BibitemShut
  {NoStop}%
\bibitem [{\citenamefont {Li}\ \emph {et~al.}(2019)\citenamefont {Li},
  \citenamefont {Liu}, \citenamefont {Huo}, \citenamefont {Cui}, \citenamefont
  {Feng}, \citenamefont {Ou},\ and\ \citenamefont {Li}}]{Li2019OE}%
  \BibitemOpen
  \bibfield  {author} {\bibinfo {author} {\bibfnamefont {J.}~\bibnamefont
  {Li}}, \bibinfo {author} {\bibfnamefont {Y.}~\bibnamefont {Liu}}, \bibinfo
  {author} {\bibfnamefont {N.}~\bibnamefont {Huo}}, \bibinfo {author}
  {\bibfnamefont {L.}~\bibnamefont {Cui}}, \bibinfo {author} {\bibfnamefont
  {C.}~\bibnamefont {Feng}}, \bibinfo {author} {\bibfnamefont {Z.~Y.}\
  \bibnamefont {Ou}},\ and\ \bibinfo {author} {\bibfnamefont {X.}~\bibnamefont
  {Li}},\ }\bibfield  {title} {\enquote {\bibinfo {title} {Pulsed entanglement
  measured by parametric amplifier assisted homodyne detection},}\ }\href@noop
  {} {\bibfield  {journal} {\bibinfo  {journal} {Opt. Express}\ }\textbf
  {\bibinfo {volume} {27}},\ \bibinfo {pages} {30552--30562} (\bibinfo {year}
  {2019})}\BibitemShut {NoStop}%
\bibitem [{\citenamefont {Shelby}\ \emph {et~al.}(1986)\citenamefont {Shelby},
  \citenamefont {Levenson}, \citenamefont {Perlmutter}, \citenamefont {DeVoe},\
  and\ \citenamefont {Walls}}]{Shelby1986PRL}%
  \BibitemOpen
  \bibfield  {author} {\bibinfo {author} {\bibfnamefont {R.~M.}\ \bibnamefont
  {Shelby}}, \bibinfo {author} {\bibfnamefont {M.~D.}\ \bibnamefont
  {Levenson}}, \bibinfo {author} {\bibfnamefont {S.~H.}\ \bibnamefont
  {Perlmutter}}, \bibinfo {author} {\bibfnamefont {R.~G.}\ \bibnamefont
  {DeVoe}},\ and\ \bibinfo {author} {\bibfnamefont {D.~F.}\ \bibnamefont
  {Walls}},\ }\bibfield  {title} {\enquote {\bibinfo {title} {Broad-band
  parametric deamplification of quantum noise in an optical fiber},}\
  }\href@noop {} {\bibfield  {journal} {\bibinfo  {journal} {Phys. Rev. Lett.}\
  }\textbf {\bibinfo {volume} {57}},\ \bibinfo {pages} {691--694} (\bibinfo
  {year} {1986})}\BibitemShut {NoStop}%
\bibitem [{\citenamefont {Yoshikawa}\ \emph {et~al.}(2016)\citenamefont
  {Yoshikawa}, \citenamefont {Yokoyama}, \citenamefont {Kaji}, \citenamefont
  {Sornphiphatphong}, \citenamefont {Shiozawa}, \citenamefont {Makino},\ and\
  \citenamefont {Furusawa}}]{Yoshikawa2016APLPhotonics}%
  \BibitemOpen
  \bibfield  {author} {\bibinfo {author} {\bibfnamefont {J.-i.}\ \bibnamefont
  {Yoshikawa}}, \bibinfo {author} {\bibfnamefont {S.}~\bibnamefont {Yokoyama}},
  \bibinfo {author} {\bibfnamefont {T.}~\bibnamefont {Kaji}}, \bibinfo {author}
  {\bibfnamefont {C.}~\bibnamefont {Sornphiphatphong}}, \bibinfo {author}
  {\bibfnamefont {Y.}~\bibnamefont {Shiozawa}}, \bibinfo {author}
  {\bibfnamefont {K.}~\bibnamefont {Makino}},\ and\ \bibinfo {author}
  {\bibfnamefont {A.}~\bibnamefont {Furusawa}},\ }\bibfield  {title} {\enquote
  {\bibinfo {title} {Invited article: Generation of one-million-mode
  continuous-variable cluster state by unlimited time-domain multiplexing},}\
  }\href@noop {} {\bibfield  {journal} {\bibinfo  {journal} {APL Photonics}\
  }\textbf {\bibinfo {volume} {1}},\ \bibinfo {pages} {060801} (\bibinfo {year}
  {2016})}\BibitemShut {NoStop}%
\bibitem [{\citenamefont {Asavanant}\ \emph {et~al.}(2019)\citenamefont
  {Asavanant}, \citenamefont {Shiozawa}, \citenamefont {Yokoyama},
  \citenamefont {Charoensombutamon}, \citenamefont {Emura}, \citenamefont
  {Alexander}, \citenamefont {Takeda}, \citenamefont {ichi Yoshikawa},
  \citenamefont {Menicucci}, \citenamefont {Yonezawa},\ and\ \citenamefont
  {Furusawa}}]{Asavanant2019science}%
  \BibitemOpen
  \bibfield  {author} {\bibinfo {author} {\bibfnamefont {W.}~\bibnamefont
  {Asavanant}}, \bibinfo {author} {\bibfnamefont {Y.}~\bibnamefont {Shiozawa}},
  \bibinfo {author} {\bibfnamefont {S.}~\bibnamefont {Yokoyama}}, \bibinfo
  {author} {\bibfnamefont {B.}~\bibnamefont {Charoensombutamon}}, \bibinfo
  {author} {\bibfnamefont {H.}~\bibnamefont {Emura}}, \bibinfo {author}
  {\bibfnamefont {R.~N.}\ \bibnamefont {Alexander}}, \bibinfo {author}
  {\bibfnamefont {S.}~\bibnamefont {Takeda}}, \bibinfo {author} {\bibfnamefont
  {J.}~\bibnamefont {ichi Yoshikawa}}, \bibinfo {author} {\bibfnamefont
  {N.~C.}\ \bibnamefont {Menicucci}}, \bibinfo {author} {\bibfnamefont
  {H.}~\bibnamefont {Yonezawa}},\ and\ \bibinfo {author} {\bibfnamefont
  {A.}~\bibnamefont {Furusawa}},\ }\bibfield  {title} {\enquote {\bibinfo
  {title} {Generation of time-domain-multiplexed two-dimensional cluster
  state},}\ }\href@noop {} {\bibfield  {journal} {\bibinfo  {journal}
  {Science}\ }\textbf {\bibinfo {volume} {366}},\ \bibinfo {pages} {373--376}
  (\bibinfo {year} {2019})}\BibitemShut {NoStop}%
\bibitem [{\citenamefont {Larsen}\ \emph
  {et~al.}(2019{\natexlab{a}})\citenamefont {Larsen}, \citenamefont {Guo},
  \citenamefont {Breum}, \citenamefont {Neergaard-Nielsen},\ and\ \citenamefont
  {Andersen}}]{Larsen2019science}%
  \BibitemOpen
  \bibfield  {author} {\bibinfo {author} {\bibfnamefont {M.~V.}\ \bibnamefont
  {Larsen}}, \bibinfo {author} {\bibfnamefont {X.}~\bibnamefont {Guo}},
  \bibinfo {author} {\bibfnamefont {C.~R.}\ \bibnamefont {Breum}}, \bibinfo
  {author} {\bibfnamefont {J.~S.}\ \bibnamefont {Neergaard-Nielsen}},\ and\
  \bibinfo {author} {\bibfnamefont {U.~L.}\ \bibnamefont {Andersen}},\
  }\bibfield  {title} {\enquote {\bibinfo {title} {Deterministic generation of
  a two-dimensional cluster state},}\ }\href@noop {} {\bibfield  {journal}
  {\bibinfo  {journal} {Science}\ }\textbf {\bibinfo {volume} {366}},\ \bibinfo
  {pages} {369--372} (\bibinfo {year} {2019}{\natexlab{a}})}\BibitemShut
  {NoStop}%
\bibitem [{\citenamefont {Huo}\ \emph {et~al.}(2020)\citenamefont {Huo},
  \citenamefont {Liu}, \citenamefont {Li}, \citenamefont {Cui}, \citenamefont
  {Chen}, \citenamefont {Palivela}, \citenamefont {Xie}, \citenamefont {Li},\
  and\ \citenamefont {Ou}}]{HuoNan2020PRL}%
  \BibitemOpen
  \bibfield  {author} {\bibinfo {author} {\bibfnamefont {N.}~\bibnamefont
  {Huo}}, \bibinfo {author} {\bibfnamefont {Y.}~\bibnamefont {Liu}}, \bibinfo
  {author} {\bibfnamefont {J.}~\bibnamefont {Li}}, \bibinfo {author}
  {\bibfnamefont {L.}~\bibnamefont {Cui}}, \bibinfo {author} {\bibfnamefont
  {X.}~\bibnamefont {Chen}}, \bibinfo {author} {\bibfnamefont {R.}~\bibnamefont
  {Palivela}}, \bibinfo {author} {\bibfnamefont {T.}~\bibnamefont {Xie}},
  \bibinfo {author} {\bibfnamefont {X.}~\bibnamefont {Li}},\ and\ \bibinfo
  {author} {\bibfnamefont {Z.~Y.}\ \bibnamefont {Ou}},\ }\bibfield  {title}
  {\enquote {\bibinfo {title} {Direct temporal mode measurement for the
  characterization of temporally multiplexed high dimensional quantum
  entanglement in continuous variables},}\ }\href@noop {} {\bibfield  {journal}
  {\bibinfo  {journal} {Phys. Rev. Lett.}\ }\textbf {\bibinfo {volume} {124}},\
  \bibinfo {pages} {213603} (\bibinfo {year} {2020})}\BibitemShut {NoStop}%
\bibitem [{\citenamefont {Shinjo}, \citenamefont {Eto},\ and\ \citenamefont
  {Hirano}(2019)}]{shinjo2019pulse}%
  \BibitemOpen
  \bibfield  {author} {\bibinfo {author} {\bibfnamefont {A.}~\bibnamefont
  {Shinjo}}, \bibinfo {author} {\bibfnamefont {Y.}~\bibnamefont {Eto}},\ and\
  \bibinfo {author} {\bibfnamefont {T.}~\bibnamefont {Hirano}},\ }\bibfield
  {title} {\enquote {\bibinfo {title} {Pulse-resolved measurement of
  continuous-variable einstein-podolsky-rosen entanglement with shaped local
  oscillators},}\ }\href@noop {} {\bibfield  {journal} {\bibinfo  {journal}
  {Optics Express}\ }\textbf {\bibinfo {volume} {27}},\ \bibinfo {pages}
  {17610--17619} (\bibinfo {year} {2019})}\BibitemShut {NoStop}%
\bibitem [{\citenamefont {Okubo}\ \emph {et~al.}(2008)\citenamefont {Okubo},
  \citenamefont {Hirano}, \citenamefont {Zhang},\ and\ \citenamefont
  {Hirano}}]{OkuboOL08}%
  \BibitemOpen
  \bibfield  {author} {\bibinfo {author} {\bibfnamefont {R.}~\bibnamefont
  {Okubo}}, \bibinfo {author} {\bibfnamefont {M.}~\bibnamefont {Hirano}},
  \bibinfo {author} {\bibfnamefont {Y.}~\bibnamefont {Zhang}},\ and\ \bibinfo
  {author} {\bibfnamefont {T.}~\bibnamefont {Hirano}},\ }\bibfield  {title}
  {\enquote {\bibinfo {title} {Pulse-resolved measurement of quadrature phase
  amplitudes of squeezed pulse trains at a repetition rate of 76 mhz},}\
  }\href@noop {} {\bibfield  {journal} {\bibinfo  {journal} {Opt. Lett.}\
  }\textbf {\bibinfo {volume} {33}},\ \bibinfo {pages} {1458--1460} (\bibinfo
  {year} {2008})}\BibitemShut {NoStop}%
\bibitem [{\citenamefont {Ou}(2017)}]{Ou2017Book}%
  \BibitemOpen
  \bibfield  {author} {\bibinfo {author} {\bibfnamefont {Z.~Y.}\ \bibnamefont
  {Ou}},\ }\href {https://doi.org/10.1142/10453} {\emph {\bibinfo {title}
  {Quantum Optics for Experimentalists}}}\ (\bibinfo  {publisher} {WORLD
  SCIENTIFIC},\ \bibinfo {year} {2017})\BibitemShut {NoStop}%
\bibitem [{\citenamefont {Guo}\ \emph {et~al.}(2013)\citenamefont {Guo},
  \citenamefont {Li}, \citenamefont {Liu},\ and\ \citenamefont
  {Ou}}]{guo2013PRA}%
  \BibitemOpen
  \bibfield  {author} {\bibinfo {author} {\bibfnamefont {X.}~\bibnamefont
  {Guo}}, \bibinfo {author} {\bibfnamefont {X.}~\bibnamefont {Li}}, \bibinfo
  {author} {\bibfnamefont {N.}~\bibnamefont {Liu}},\ and\ \bibinfo {author}
  {\bibfnamefont {Z.~Y.}\ \bibnamefont {Ou}},\ }\bibfield  {title} {\enquote
  {\bibinfo {title} {Multimode theory of pulsed-twin-beam generation using a
  high-gain fiber-optical parametric amplifier},}\ }\href@noop {} {\bibfield
  {journal} {\bibinfo  {journal} {Physical Review A}\ }\textbf {\bibinfo
  {volume} {88}},\ \bibinfo {pages} {023841} (\bibinfo {year}
  {2013})}\BibitemShut {NoStop}%
\bibitem [{\citenamefont {Du}\ \emph {et~al.}(2018)\citenamefont {Du},
  \citenamefont {Li}, \citenamefont {Liu}, \citenamefont {Wang},\ and\
  \citenamefont {Li}}]{LiYongmin2018JOSAB}%
  \BibitemOpen
  \bibfield  {author} {\bibinfo {author} {\bibfnamefont {S.}~\bibnamefont
  {Du}}, \bibinfo {author} {\bibfnamefont {Z.}~\bibnamefont {Li}}, \bibinfo
  {author} {\bibfnamefont {W.}~\bibnamefont {Liu}}, \bibinfo {author}
  {\bibfnamefont {X.}~\bibnamefont {Wang}},\ and\ \bibinfo {author}
  {\bibfnamefont {Y.}~\bibnamefont {Li}},\ }\bibfield  {title} {\enquote
  {\bibinfo {title} {High-speed time-domain balanced homodyne detector for
  nanosecond optical field applications},}\ }\href@noop {} {\bibfield
  {journal} {\bibinfo  {journal} {J. Opt. Soc. Am. B}\ }\textbf {\bibinfo
  {volume} {35}},\ \bibinfo {pages} {481--486} (\bibinfo {year}
  {2018})}\BibitemShut {NoStop}%
\bibitem [{\citenamefont {Larsen}\ \emph
  {et~al.}(2019{\natexlab{b}})\citenamefont {Larsen}, \citenamefont {Guo},
  \citenamefont {Breum}, \citenamefont {Neergaard-Nielsen},\ and\ \citenamefont
  {Andersen}}]{Larsen2019njpQI}%
  \BibitemOpen
  \bibfield  {author} {\bibinfo {author} {\bibfnamefont {M.~V.}\ \bibnamefont
  {Larsen}}, \bibinfo {author} {\bibfnamefont {X.}~\bibnamefont {Guo}},
  \bibinfo {author} {\bibfnamefont {C.~R.}\ \bibnamefont {Breum}}, \bibinfo
  {author} {\bibfnamefont {J.~S.}\ \bibnamefont {Neergaard-Nielsen}},\ and\
  \bibinfo {author} {\bibfnamefont {U.~L.}\ \bibnamefont {Andersen}},\
  }\bibfield  {title} {\enquote {\bibinfo {title} {Fiber-coupled epr-state
  generation using a single temporally multiplexed squeezed light source},}\
  }\href@noop {} {\bibfield  {journal} {\bibinfo  {journal} {npj Quantum
  Information}\ }\textbf {\bibinfo {volume} {5}},\ \bibinfo {pages} {46}
  (\bibinfo {year} {2019}{\natexlab{b}})}\BibitemShut {NoStop}%
\bibitem [{\citenamefont {Dong}\ \emph {et~al.}(2017)\citenamefont {Dong},
  \citenamefont {Yao}, \citenamefont {Zhang}, \citenamefont {Chen},
  \citenamefont {Zhang}, \citenamefont {You}, \citenamefont {Wang},\ and\
  \citenamefont {Huang}}]{2K_DSF}%
  \BibitemOpen
  \bibfield  {author} {\bibinfo {author} {\bibfnamefont {S.}~\bibnamefont
  {Dong}}, \bibinfo {author} {\bibfnamefont {X.}~\bibnamefont {Yao}}, \bibinfo
  {author} {\bibfnamefont {W.}~\bibnamefont {Zhang}}, \bibinfo {author}
  {\bibfnamefont {S.}~\bibnamefont {Chen}}, \bibinfo {author} {\bibfnamefont
  {W.}~\bibnamefont {Zhang}}, \bibinfo {author} {\bibfnamefont
  {L.}~\bibnamefont {You}}, \bibinfo {author} {\bibfnamefont {Z.}~\bibnamefont
  {Wang}},\ and\ \bibinfo {author} {\bibfnamefont {Y.}~\bibnamefont {Huang}},\
  }\bibfield  {title} {\enquote {\bibinfo {title} {True single-photon
  stimulated four-wave mixing},}\ }\href@noop {} {\bibfield  {journal}
  {\bibinfo  {journal} {ACS Photonics}\ }\textbf {\bibinfo {volume} {4}},\
  \bibinfo {pages} {746--753} (\bibinfo {year} {2017})}\BibitemShut {NoStop}%
\end{thebibliography}%

\end{document}